\documentclass{vgtc}                          % final (conference style)
\graphicspath{{figures/}{pictures/}{images/}{./}} % where to search for the images

\usepackage{times}                     % we use Times as the main font
\usepackage{float}
\usepackage{tabu}  
\usepackage{enumitem}% only used for the table example
\usepackage{booktabs}                  % only used for the table example
\usepackage{lipsum}                    % used to generate placeholder text
\usepackage{mwe}                       % used to generate placeholder figures
\usepackage{listings} 
\usepackage{mathptmx}                  % use matching math font
\usepackage{amsmath}
\usepackage{balance}
\onlineid{8466}

\vgtccategory{Research}

\vgtcinsertpkg

\title{UltraArUco: A Lightweight Multilingual Library and Framework with Low-Latency Real-Time Marker-Based Tracking System for Mobile AR Interaction}

\author{Mikhail Kiselev$^1$ \thanks{E-mail: Mikhail.Kiselev@skoltech.ru}%
\and Aleksandr Marukhin$^1$ \thanks{E-mail: Aleksandr.Marukhin@skoltech.ru}%
\and Ivan Snegirev\thanks{E-mail: Ivan.Snegirev@skoltech.ru} %
\and Elizaveta Semenyakina\thanks{E-mail: Elizaveta.Semenyakina@skoltech.ru} %
\and Miguel Altamirano Cabrera\thanks{E-mail: m.altamirano@skoltech.ru} %
\and Dzmitry Tsetserukou\thanks{E-mail: d.tsetserukov@skoltech.ru}}
\affiliation{\scriptsize Skolkovo Institute of Science and Technology, Russian Federation\\
 $^1$ These authors contributed equally to this work}%\author{Anonymous Author(s)}

\teaser{
  \centering
  \hspace*{-0.8cm}\includegraphics[width=1.1\linewidth]{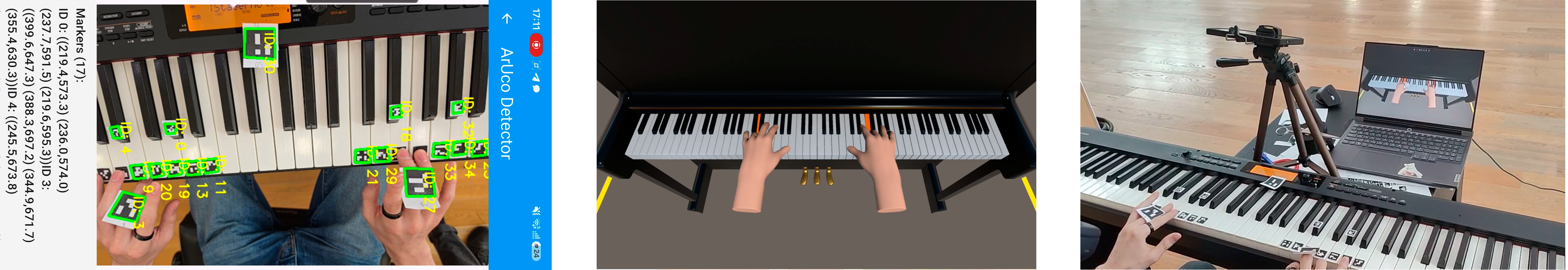}
  \caption{Demonstration of the UltraArUco application from real-time camera detection of  ArUco markers in mobile app to the piano playing visualization in Unity and real and simulated playing experience example. Note: Real piano is shown only for reference comparison. ArUco keys can be played without it.}
  \label{fig:teaser}
}

\abstract{
    UltraArUco - a lightweight multilingual library and framework for low-latency, real-time marker-based tracking in mobile augmented reality. Unlike standard OpenCV-based implementations, UltraArUco introduces an optimized multilingual wrapper that reduces per-frame latency in six times, while maintaining high accuracy. Distributed Wi-Fi architecture provides portability, connects mobile device (camera-input) with a PC-based visual application, enabling responsive interactions. The framework is validated through an interactive piano simulation, where static ArUco markers on keys enable occlusion-based note triggering, and hand-mounted markers provide spatial gesture recognition. UltraArUco’s system requirements make it highly suitable for resource-constrained mobile AR-applications, demonstrating a viable AR music application without specialized equipment.

} % end of abstract


\begin{document}


\firstsection{Introduction}

\maketitle

Marker-based augmented reality (AR) has become a valuable part of spatial computing, allowing precise camera pose estimation and interaction in a variety of applications, from robotics and education to the household routines~\cite{articleGarrido}. Among fiducial marker systems, ArUco~\cite{articleGarrido} has emerged as a popular solution due to its seamless integration with OpenCV and comparably fast and precise marker detection. However, the standard OpenCV implementation relies on general-purpose computer vision algorithms that carry computational overhead not needed for simple square fiducial marker detection, especially ArUco. That results in significant computational bottlenecks—particularly on resource-constrained mobile devices where fast real-time performance is critical.

Recent advances have introduced ArUco Nano~\cite{GARRIDOJURADO2026102690}, a minimalist header-only C++ pure algorithm that achieved up to 6.5× speedup over OpenCV through novel algorithmic additions, including visited-aware contour extraction and direct sub-pixel code sampling. However, it lacks native bindings  for modern frameworks, limiting rapid AR prototyping. Furthermore, it has not been validated to verify its declared recognition speed.

In this work, we introduce UltraArUco, a lightweight multilingual framework that bridges this gap. We validate its real-world performance through our novel AR piano application, demonstrating its superior low-latency tracking over standard OpenCV implementations in latency-critical musical interactions. Materials are available on GitHub: https://github.com/Alexander-ha/-ULTRAARUCO-ArucoPiano . The demonstration of the application is shown in Fig.~\ref{fig:teaser}. Key insights of our work:

\textbf{Multilingual FFI Wrapper:} This is the core contribution of our work. We provide a C-based dynamic library interface that enables seamless integration with Flutter, Python, C\#, Java and Node.js by Foreign Function Interface (FFI), preserving the performance advantages of ArUco Nano with custom optimization while enabling cross-platform development.

%The core contribution of our work is the Multilingual FFI Wrapper: We provide a C-based dynamic library interface that enables seamless integration with Flutter, Python, C#, Java, and Node.js by Foreign Function Interface (FFI). This wrapper successfully preserves the performance advantages of the ArUco Nano core while enabling cross-platform mobile development. To demonstrate and validate this contribution, we present two applications:
    
\textbf{Distributed AR Architecture:} A Wi-Fi-based pipeline connecting a mobile device (handling camera capture and marker detection) with a PC-based Unity application (3D rendering and visualization), achieving low-latency real-time tracking suitable for immersive entertainment experiences.
    
\textbf{Interactive Piano Demonstration:} An AR music application using ArUco markers on piano keys and hands for occlusion-based gesture recognition and hand-position tracking, enabling AR music interaction without specialized equipment; a piano was used only to validate spatial alignment between real and visualized notes.

\section{Related Work}

\subsection{Fiducial Marker Systems}
ArUco~\cite{articleGarrido} became standard for AR/robotics pose estimation due to configurable, occlusion-robust marker dictionaries. AprilTags~\cite{iturralde} offer similar capabilities with better noise resistance but slower recognition, while ArUco Nano~\cite{GARRIDOJURADO2026102690} achieved a 6.5× speedup by optimizing tracing and local-mean computation, avoiding costly contour extraction and warping.

\subsection{Mobile AR Frameworks and AR in Music Interaction }
ARCore (Google) and ARKit (Apple) provide markerless tracking through SLAM and plane detection. However, they lack the precision of fiducial markers for fine-grained interaction. Open-source alternatives like OpenCV's ArUco module offer flexibility but suffer from performance limitations on mobile devices, typically achieving 10-15 FPS or even lower on mid-range powerful smartphones for just a couple of markers, compared to 30+ FPS on ten or more markers required for smooth interaction.

AR music applications~\cite{arMusicEducation} typically rely on specialized hardware or markerless tracking that struggles with rapid finger movements. Marker-based systems offer a lightweight alternative but suffer from high latency - a challenge that UltraArUco resolves.

%Also, augmented reality has been explored for music applications, with systems that project visual guides onto physical instruments~\cite{arMusicEducation}. However, most approaches require specialized hardware (depth cameras, motion capture) or markerless tracking that struggles with the rapid finger movements in piano performance. Marker-based systems offer a lightweight alternative but have not been widely adopted due to latency concerns—precisely the challenge that UltraArUco addresses.

\section{System Architecture}

\begin{figure}[t]
    \centering
    \includegraphics[width=\linewidth]{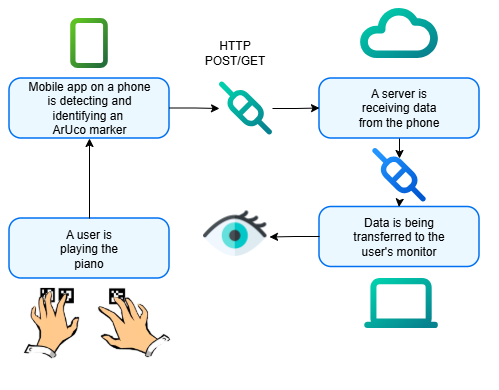}
    \caption{The diagram of the interaction pipeline.}
    \label{fig:schematic}
    \vspace{-0.75cm}
\end{figure}

UltraArUco consists of printed ArUco markers, a mobile app with an optimized multilingual detection library, and a Unity 3D visualization app. The mobile device performs real-time marker tracking and occlusion, streaming spatial coordinates over a local network to Unity for visualization. Its key novelty is the lightweight UltraArUco library, which handles all computations on-device, while Unity remains a visualization, validation, and interaction layer.
Detailed pipeline is presented in Fig. ~\ref{fig:schematic}.

\vspace{-0.25cm}
\subsection{UltraArUco Library}
UltraArUco is implemented as a C-wrapper around the ArUco Nano C++ core, exposing a minimal  Application Programming Interface (API) through a dynamic shared library (.so/.dll/.dylib):
\vspace{-0.25cm}
\begin{lstlisting}[language=C, caption={UltraArUco C API}]
ArucoDetectorHandle create_detector();
void destroy_detector(ArucoDetectorHandle handle);
int detect_markers(
    ArucoDetectorHandle handle,
    const uint8_t* data, int width, int height,
    int* ids, float* corners, int max_markers,
    float* processing_time_ms
);
\end{lstlisting}
ArucoDetectorHandle is a type alias for aruco\_nano::ArucoDetector's classical type, which can be accessed and configured with a wide range of parameters, defined by the user. It preserves flexibility up to redefining each argument in the class, while the core methods deliver minimal and simple API.
Here, the core methods are presented:
\begin{itemize}[nosep]
    \item \textbf{create$\_$detector()} is for creating a new object with the detector's type.
    \item \textbf{destroy$\_$detector()} is a destructor of an object.
    \item \textbf{detect$\_$markers()} is the method that returns the integer tag of a marker and receives as an input type alias, a pointer to the image data, the width, and height of the image. It also updates indices and a float array of corners for the specific marker. max$\_$markers parameter controls the number of detectable markers per frame, and the last argument stands for measuring the performance.
\end{itemize}

This design enables integration with any language that supports FFI. For example, in Dart/Flutter it can be demonstrated through creating type alias and then the function based on this alias:
\vspace{-0.25cm}
\begin{lstlisting}[language=Java, caption={Dart FFI Bindings}]
typedef DetectMarkersNative = Int32 Function(
    Pointer<Void> handle, Pointer<Uint8> data,
    Int32 width, Int32 height,
    Pointer<Int32> ids, Pointer<Float> corners,
    Int32 maxMarkers, Pointer<Float> processingTime);
final detectMarkers = nativeLib
    .lookup<NativeFunction<DetectMarkersNative>>('detect_markers')
    .asFunction<DetectMarkersDart>();
\end{lstlisting}

The library preserves ArUco Nano's core optimizations: tolerance threshold $R_{max}$ for noise rejection, box filter ($S_{min}$) for local mean computation (1.2× speedup over adaptive thresholding), and Visited-Aware contour extraction avoiding redundant node revisits.

\subsection{Compilation Optimizations}
We evaluated compiler flags to maximize detection throughput. 
The $-\text{O}0$ implies no optimization,  while $-\text{O}2$ utilizes aggressive optimization such as replacing expensive CPU-operations with faster alternatives and assigning frequently used variables to CPU registers. Additionally, we evaluated  $-\text{O}3$, which introduces loop unrolling and vectorization, and $-\text{O}fast$, which prioritizes execution speed over strict mathematical compliance. %Table~\ref{tab:compiler_flags} shows that $-\text{O}2$ achieves the best balance, providing a 3.06× speedup over the baseline $-\text{O}0$ while maintaining numerical stability.
Among the evaluated configurations (see Table~\ref{tab:compiler_flags}), $-\text{O}2$ yielded the lowest execution time, reducing the mean processing time from 329.21 ms to 107.68 ms, corresponding to a 3.06× speedup over the unoptimized $-\text{O}0$ build.
 
\begin{table}[htbp]
\centering
\caption{Compiler Flags Performance Comparison.}
\label{tab:compiler_flags}
\small
\begin{tabular}{p{4.5cm} c c}
\toprule
Compiler Flags & Time (ms) & Speedup vs O0 \\
\midrule
-O2 & 107.68 & 3.06× \\
-O3 -funroll-loops -ftree-vectorize & 125.95 & 2.61× \\
-Ofast -ffast-math & 142.38 & 2.31× \\
-O0 (baseline) & 329.21 & 1.00× \\
\bottomrule
\end{tabular}
 \vspace{-0.5cm}
\end{table}

The test was performed on the original static image dataset for the inference of ArUco Nano  \cite{GARRIDOJURADO2026102690} at high resolution (24 MP) of the samples. The optimal configuration in our application is chosen based on the table above. All compilations were performed with CMake.

\subsection{Mobile Detection Application}
The mobile app is built in Flutter with the following pipeline:
\begin{enumerate}[nosep]
    \item \textbf{Camera Capture:} 60+ FPS stream at 1280×720 resolution
    \item \textbf{Frame Preprocessing:} Color space conversion (YUV→ RGBA), rotation compensation based on device orientation
    \item \textbf{Marker Detection:} UltraArUco processes markers' set in approximately 5 ms in our application, detecting up to 40 markers simultaneously (theoretically could be increased to $\simeq$70)
    \item \textbf{Data Transmission:} Detected marker IDs and 2D corner coordinates are serialized as JSON and transmitted via HTTP POST to the Unity server at 30 Hz.
\end{enumerate}

\subsection{Wi-Fi Communication Protocol}
The mobile device and PC communicate over local Wi-Fi using a lightweight HTTP-based protocol. Each frame transmits a JSON payload:
\vspace{-0.35cm}
\begin{lstlisting}[caption={Marker Data Payload}]
{
  "frame_id": 1234,
  "timestamp": 1624567890.123,
  "markers": [
    {"id": 1, "corners": [[x1,y1], [x2,y2], [x3,y3], [x4,y4]]},
    {"id": 2, "corners": [...]}
  ]
}
\end{lstlisting}
End-to-end latency (capture → detection → Wi-fi transmission → Unity rendering) averages 35 ms, well within the 100 ms threshold for perceived real-time interaction~\cite{latencyThreshold}.

\subsection{Visualization Pipeline}
% The application receives marker data and reconstructs the 3D scene:
The Unity application receives the detected marker IDs and image-plane corner coordinates from the mobile device. Unity maps these coordinates to a calibrated keyboard coordinate system and updates the positions of the virtual keyboard and hand representations. The markers are categorized into three functional groups:

% \textbf{Marker Roles:} Markers are categorized into three groups:
\begin{itemize}[nosep]
    \item \textbf{Piano Anchor (1 marker):} Defines the reference position and orientation of the keyboard within the calibrated interaction area. Its detected corner coordinates are used to align the virtual keyboard with the keyboard coordinate system.
    \item \textbf{Hand Trackers (2 markers):} Mounted on the user’s hands, these markers provide hand-position information. Each marker is mapped from image-plane coordinates to the calibrated keyboard coordinate system using hand-specific calibration parameters.
    \item \textbf{Musical Key Markers (18 markers):} One marker per playable piano key. Occlusion of a key marker triggers the corresponding virtual key animation and note playback.
    \\\ Note: The number of key markers can be increased, but that amount was noticed as enough for the experiment. 
\end{itemize}

\textbf{Interaction Logic:} One controller maps markers from image to world space, positioning the piano in a calibrated zone with adjustable parameters. Hand tracking utilizes per-hand calibration and an Ultraleap "Ghost Hands" visualization rig driven by axis-free Cyclic Coordinate Descent (CCD) inverse kinematics, employing an order-preserving algorithm to prevent finger crossing during chords.

\textbf{Note Triggering:} Keys depress upon finger contact, triggering hinge animations, emissive glow, particle effects, and audio feedback. The data-source-agnostic pipeline is fully configurable via Unity inspector parameters, enabling scene recalibration without code modifications.
\section{Experiments}

\subsection{Experimental Setup}
To validate our lightweight multilingual framework, we conducted experiments with our AR piano application, using different parameters.
Different hardware and configurations for UltraArUco validation:
\begin{itemize} [nosep]
    \item \textbf{Mobile Device:} Samsung Galaxy S26 Ultra (Snapdragon Elite 8 Gen 5, 12GB RAM, Android 16), Xiaomi Redmi Note 11 Pro (MediaTek Helio G96, 8GB RAM, Android 13)
    \item \textbf{PC:} Intel Core i7-12450H, 16GB RAM, Windows 11
    \item \textbf{Network:} Wi-Fi 6 (802.11ax), 5GHz band, $<10$ ms round-trip latency
    \item \textbf{Markers:} 4×4 dictionary, 21 unique IDs, printed in different size on matte paper for different purposes (hands, piano, or musical keys)
    \item \textbf{Piano:} 88-key Casio piano, 19 ArUco markers: 18 affixed to keys of closely two octaves,1 as a piano anchor. 
\end{itemize}

\subsection{Performance Benchmarks}
Table~\ref{tab:detection_performance} compares UltraArUco against OpenCV's standard ArUco and ArUco Nano implementation across multiple resolutions.
\vspace{-0.3 cm}
\begin{table}[h!]
\centering
\caption{Detection performance comparison for one marker and the speedup ratio of UltraArUco to OpenCV (obtained on Samsung Galaxy S26 Ultra).}
\label{tab:detection_performance}
\scriptsize
\setlength{\tabcolsep}{1pt}  
\begin{tabular}{l|c|c|c|c}
\toprule
Resolution & UltraArUco & OpenCV (1-thread) & ArUco Nano & Speedup\\
\midrule
640×360 (0.2 MP) & 1.61 ms& 11.61 ms& 3.9 ms& 7.2×\\
1280×720 (0.9 MP) & 3.0 ms& 20.1 ms& 7.8 ms& 6.7×\\
1920×1080 (2.1 MP) & 11.2 ms& 60.5 ms& 30.2 ms& 5.4×\\
3840×2160 (8.3 MP) & 42.7 ms& 210.5 ms& 109.3 ms& 4.9×\\
\bottomrule
\end{tabular}
\end{table}
\vspace{-0.3cm}

\begin{figure}[t]
    \centering
    % \begin{subfigure}{\columnwidth}
        \centering
        \includegraphics[width=\linewidth, trim=0 0 0 16, clip]{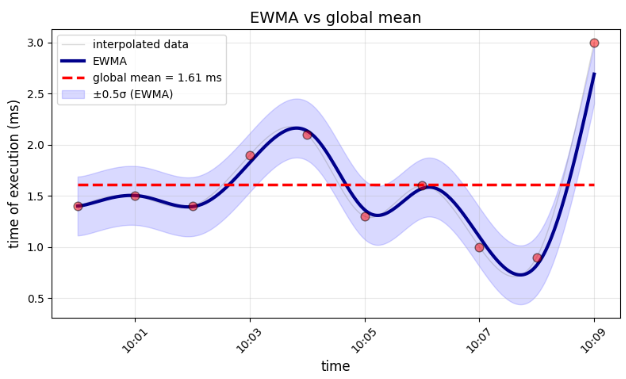}
        % \vspace{-0.5cm}
        \caption{EWMA and global mean of UltraAruco implementation of ArUcoNano method's  performance (ms/1 detection).}
        
        \label{fig:a}
\end{figure}

\begin{figure}[t]
        \centering
        \includegraphics[width=\linewidth, trim=0 0 0 11, clip]{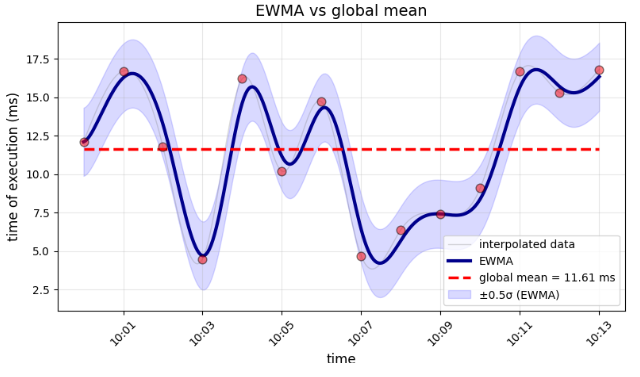}
        \caption{EWMA and global mean of OpenCV method's  performance (ms/1 detection).}
        \label{fig:b}
        \vspace{-0.5cm}
\end{figure}

At the operational resolution of 1280×720, UltraArUco achieves 3.0 ms per frame, enabling 30+ FPS detection with substantial headroom for preprocessing and transmission. This represents a 6.7× speedup over OpenCV for 1 marker detection, comparable with ArUco Nano's reported theoretical performance~\cite{GARRIDOJURADO2026102690}.\\
Speed-up measurements were performed with the following methodology:\\
1. The results obtained on the same marker from both UltraArUco and OpenCV methods were collected during 5 minutes of continuous monitoring.\\
2. EWMA (Exponentially Weighted Moving Average) \cite{ewma} was calculated on the measured data using the following expression: $\mu_t = \lambda X_t + (1-\lambda)\mu_{t-1}$, where $X_t$ is the measured data at moment t, $\lambda = 0.2$ is the threshold and $\mu_{t-1}$ is the metric from the previous step. The control boundaries are represented as follows $\mu_t \pm L \cdot \sigma \sqrt{\frac{\lambda}{2-\lambda} \left( 1 - (1-\lambda)^{2t} \right)}$.\\
3. The global mean of the measured data was added to Table~\ref{tab:detection_performance}.\\
EWMA was selected for performance estimation to appropriately account for thermal throttling that occurs over continuous operation.
Most of the experiments were performed on Snapdragon Elite 8 Gen 5, but for compatibility tests, some of them were also performed for compatibility on MediaTek Helio G96.
The difference of performance's stability (monitoring for $>5$ minutes) of both methods is demonstrated on Fig. ~\ref{fig:a} and Fig. ~\ref{fig:b} .

    \vspace{1em}
    
%     \caption{Comparison of performance's stability (monitoring for $>5$ minutes) on Snapdragon 8 Gen 3.}
%     \label{fig:bigfigure}
%      \vspace{-0.65cm}
% \end{figure}
According to Fig. ~\ref{fig:a} and Fig. ~\ref{fig:b}, we can admit that our implementation is not only much faster than  OpenCV, but also demonstrates better stability with much less varying intervals of standard deviation ($\pm 0.5\sigma$). \\
The tests also showed that despite the CPU's temperature remaining similar for both tests (~40°C), the available RAM is enlarged up to 250 MB with UltraArUco implementation.

\subsection{Piano Application Results}
The interactive virtual piano is a standard benchmark for gesture recognition \cite{handTrackingPiano}. While MediaPipe is widely used \cite{goyalplay}, we propose ArUco markers for a faster and more accurate response.
We conducted a preliminary proof-of-concept demonstration with two participants (an expert and a novice), focusing on real-time single-note and chord interactions.
%%\begin{figure}[h]
    %%\centering
    %%\includegraphics[width=\linewidth]{handplay.png}
    %%\caption{Demonstration of the functionality based on the %%markers instead of real keys.}
%%    \label{fig:placeholder}
%%\end{figure}
Digital instruments aim for latency under 10 ms \cite{effectoflatency}; our mobile app achieves $1.7-3.0$ ms. The system demonstrated robust tracking, accurately detecting single and simultaneous key presses and reliably supporting continuous play for both users, translating key presses  into visual and audio feedback.
%%\begin{figure}[h]
%%    \centering
%%    \includegraphics[width=\linewidth]{foreplay.png}
%%    \caption{Demonstration of the simultaneous detection for data transferring into VR.}
%%    \label{fig:placeholder}
%%\end{figure}
False positives (around 7\% of interactions) primarily resulted from accidental marker occlusion during rapid hand transitions or complex chords. Despite this, the system was perceived as intuitive and effective for its validation and entertainment purposes.

\textbf{Limitations:} The system requires line-of-sight and suffers tracking loss at hand speeds ($>0.5$ m/s), with wired connections recommended for low latency. The 21-marker setup limits the range to closely two octaves, while extending to 88 keys increases power demands and may slow detection.

\section{Conclusion and Future Work}

UltraArUco demonstrates that optimized fiducial marker detection enables responsive AR on commodity mobile hardware. By combining ArUco Nano's optimizations with a multilingual FFI wrapper and distributed Wi-Fi architecture, we achieved 6× faster detection than OpenCV while maintaining playable end-to-end latency.

The piano application validates this framework for AR music and rapid-recognition tasks, requiring only a smartphone and printed markers. Furthermore, the occlusion-based interaction paradigm proves intuitive and robust, achieving 93\% detection accuracy.

%UltraArUco demonstrates that optimized fiducial marker detection can enable responsive, immersive AR experiences on commodity mobile hardware. By combining ArUco Nano's algorithmic innovations with a multilingual FFI wrapper and distributed Wi-Fi architecture, in our application we achieved 5× faster detection than OpenCV while maintaining sub-50 ms end-to-end latency suitable for real-time interaction.

%The piano application validates fast response of the framework for AR music entertainment and other fast-recognition use cases, offering a camera-driven interface requiring only printed markers and a smartphone. The occlusion-based interaction paradigm proves intuitive and robust with 93\% detection accuracy.

Furthermore, we suggest that such high-speed marker detection methods not only enable low-latency motion capture for virtual applications \cite{mocaptech}, but also facilitate the transfer of real-time mechanics into virtual environments. This capability paves the way for next-generation VR/AR training systems in domains such as rehabilitation, surgical simulation \cite{surgerytech}, musical education \cite{arMusicEducation}, and analysis of sports performance, where immediate response is necessary.

\textbf{Future Work:}

    We plan to expand the application by adapting the framework for other domains including surgical training (marker-based instrument tracking), industrial assembly (AR-guided procedures) andbroaden the applicability of our framework to other domains.
    Regarding marker detection, we aim to optimize recognition latency and tracking accuracy to enhance detection reliability and interaction smoothness.
    
UltraArUco is released as open-source software with documentation, precompiled binaries, and example projects available.

%%\section*{Supplemental Materials}
%%\label{sec:supplemental_materials}

%%All supplemental materials are available on GitHub. These include:
%%begin{itemize}
  %%  \item Source code for the UltraArUco C library and multilingual bindings (Dart/Flutter, Python, C\#)
    %%\item Precompiled binaries for Android (ARM64)
    %%\item Unity project with the interactive piano demonstration, including 3D models, audio samples, and calibration tools
    %%\item Video demonstrations of the system in operation
%%\end{itemize}

%%\section*{Figure Credits}
%%\label{sec:figure_credits}

%All figures and tables were created by the authors. The piano 3D model is based on a public domain design. The "Ghost Hands" rig is based on Ultraleap's open-source hand tracking visualization, modified for marker-based posing.

\balance
\section*{Acknowledgements} 
Research reported in this publication was financially supported by the RSF grant No. 24-41-02039.

\end{document}